
\input phyzzx


\let\refmark=\NPrefmark 
\def\define#1#2\par{\def#1{\Ref#1{#2}\edef#1{\noexpand\refmark{#1}}}}
\def\con#1#2\noc{\let\?=\Ref\let\<=\refmark\let\Ref=\REFS
         \let\refmark=\undefined#1\let\Ref=\REFSCON#2
         \let\Ref=\?\let\refmark=\<\refsend}

{}~\hfill\vbox{\hbox{TIFR/TH/92-30}\hbox{June, 1992}}\break

\define\SZ
M. Saadi and B. Zwiebach, Ann. Phys. {\bf 192}(1989) 213;
T. Kugo, H. Kunitomo and K. Suehiro, Phys. Lett. {\bf B226} (1989) 48;
M. Kaku and J. Lykken, Phys. Rev. {\bf D3} (1988) 3067;
M. Kaku, preprint CCNY-HEP-89-6, Osaka-OU-HET 121;
C. Schubert, MIT preprint CTP 1977;
T. Kugo and K. Suehiro, Nucl. Phys.{\bf B337} (1990) 343.

\define\MMSO
B. Lian and G. Zuckerman Phys. Lett. {\bf B266}(1991) 21; S. Mukherji, S.
Mukhi and A. Sen, Phys. Lett. {\bf B266} (1991) 337; P. Bouwknegt, J.
McCarthy and K. Plich, CERN-TH 6162/91.

\define\GROSS
D. Gross, I. Klebanov and M. Newmann, Nucl. Phys. {\bf B350}(1991) 621;
U. H. Danielsson and D. Gross, PUPT-1258.

\define\KP
I. R. Klebanov and A. M. Polyakov, Mod. Phys. Lett. {\bf A6}(1991)3273.

\define\WIN
J. Avan and A. Jevicki, Phys. Lett. {\bf B266}(1991) 35; S. R. Das, A.
Dhar, G. Mandal and S. R. Wadia TIFR/TH/91-44, 57; G. Moore and N.
Seiberg, Rutgers Preprint RU-91-29; D. Minic, J. Polchinski and Z. Yang
UTTG-16-91.

\define\WIT
E. Witten, IAS preprint, IASSNS-91/51.

\define\WZ
E. Witten and B. Zwiebach, preprint IASSNS-HEP-92/4, MIT-CTP-2057.

\define\VEN
G. Veneziano, CERN-TH-6077/91; E. Smith and J. Polchinski, UTTG-07-91;
A. A. Tseytlin, Mod. Phys. Lett. {\bf A6} (1991) 1721; R. Brandenberger
and C. Vafa, Nucl. Phys. {\bf B316} (1989) 391.

\define\DVV
R. Dijkgraaf, E. Verlinde and H. Verlinde, Comm. Math. Phys. {\bf 115}
(1988) 649.

\define\ERIC
E. Verlinde, IAS preprint, IASSNS-HEP-92/5.

\define\MH
M. Henneaux, Nucl. Phys. {\bf B}(Proc. Suppl.)18A (1990) 47.

\define\MM
J. Maharana and S. Mukherji, TIFR/TH/92-01, to appear in Phys. Lett. {\bf
B}.

\define\AS
A. Sen, Tata preprint, TIFR/TH/92-03.

\title{ TARGET SPACE INTERPRETATION OF NEW MODULI IN 2D STRING THEORY}

\author{ Swapna Mahapatra, Sudipta Mukherji and Anirvan M. Sengupta\foot{
e-mail: swapna@tifrvax, mukherji@tifrvax, anirvan@tifrvax.}}
\address{ Tata Institute of Fundamental Research\break Homi Bhabha Road,
Bombay 400 005, India}

\abstract
We analyze the new states that have recently been discovered in 2D
string theory by E. Witten and B. Zwiebach. Since the Liouville direction
is uncompactified, we show that the deformations by the new
ghost number two states
generate equivalent classical solutions of the string fields.
We argue that the new ghost number one states are responsible for generating
transformations which relate such equivalent solutions. We also
discuss the possible interpretation of higher ghost number states of
those kinds.

\endpage

\centerline{\bf 1. Introduction}

Recent investigations of critical string theory in two dimensional target
spacetime have displayed a very rich and interesting structure. One of the
most unexpected features of this theory is that it contains infinite
number of degrees of freedom
(apart from a massless boson) in its spectrum. It would seem from a naive
light-cone gauge argument would indicate that, since there are no
transverse dimensions, there are no physical excitations except for the
centre-of-mass of the string. However, it turns out that besides this
massless excitation, this model contains new non-trivial operators at
discrete values of momenta and at various ghost numbers\MMSO \foot{In the
context of matrix model it appeared in Ref.\GROSS.}.
In a subsequent
analysis\WIT , it was shown that among those, the ghost number zero BRST
invariant operators are the members of a ring called the ground ring
and the
ghost number one states are associated with symmetries of the theory.
These symmetries also appeared in a different analysis of this model\KP\WIN.

Recently, Witten and Zwiebach\WZ ~carried out a detailed BRST analysis of the
theory and, after combining the left movers and the right movers, they
found more physical discrete states and symmetries than had usually been
supposed. In this letter, we concentrate on the target space
interpretations of some of these operators which we refer to as
the new operators. Specifically, we consider
the new operators of ghost number two and one and comment on the
members of higher ghost numbers.
We find, since the Liouville coordinate is uncompactified, the
deformations by the ghost number two operators of these kinds (which
corresponds to moduli)
generate different equivalent solutions of string fields. We
explicitly demonstrate this fact by taking the lowest member of the series.
Following descent equation ~\WZ ~, we show that if we add the
corresponding term in the sigma model action as perturbation, it can
be removed by suitably redefining the matter and Liouville coordinates.
In the corresponding string field theory of
this model, these deformations can be understood as pure gauge
deformations where gauge transformation parameters are singular
in momentum space ({\it i.e.} large at the target space boundaries). In this
sense, we believe the new states are trivial. We also discuss the new
ghost number one states which can naturally be thought of
as gauge transformation
parameters in string field theory. The charges corresponding to these
parameters rotate ghost number two operators into the new operators
and hence, in a
sense act trivially on the space of gauge inequivalent solutions.

Since some of our analysis closely mimic the analysis of
the marginal deformations in usual critical string theory\DVV, we start with
that example focusing on relavent issues. In section 3,
we discuss the new ghost number two and one operators  in 2D string theory.
This letter ends with discussion on higher ghost number
operators which can be identified with the antifields in the
Batalin-Vilkovisky formalism~\ERIC.
\bigskip

\centerline{\bf 2. Marginal perturbation in critical string theory}

We start with the example of 26 dimensional critical string theory action
in flat background, where,
$$
S ={1\over 8\pi}\int {\sqrt h}h^{\alpha\beta}
\eta_{\mu\nu}\partial_{\alpha}X^{\mu}
\partial_{\beta} X^{\nu}
\eqn\one
$$
This action has a set of marginal defomations of
the form $\int \lambda_{\mu\nu}
\partial_{\alpha}X^\mu\partial_{\beta}X^\nu h^{\alpha\beta}\,{\sqrt h}$, with
$\lambda_{\mu\nu}$ being the coupling constants. With this
perturbation, the action is
$$
S_{pert.} = {1\over 8\pi}\int {\sqrt h}h^{\alpha\beta}
\eta_{\mu\nu}\partial_{\alpha}X^{\mu}
\partial_{\beta} X^{\nu} + \int \lambda_{\mu\nu}
\partial_{\alpha}X^\mu\partial_{\beta}X^\nu h^{\alpha\beta}\,{\sqrt h}.
\eqn\two
$$
Naturally, depending on whether $X^\mu$ is compact or non-compact, the
perturbed theory will behave in different
manner.

(a) $X^\mu$ uncompactified:

Redefining $X^\mu$ as
$$\eqalign{
{X^\prime}^\mu = ({{\sqrt {1 + \lambda}}})_{\mu\nu}\,X^\nu & \cr
= X_{\mu} + 1/2 \lambda_{\mu\nu}\,X^{\nu} + {\cal O}\,(\lambda^2) \cr}
\eqn\three
$$
the perturbed action reduces to the original form \one. Hence the
partition function and the correlation functions will remain unchanged
under such perturbation as long as $X^\mu$ is un-compactified.

(b) $X^\mu$ compactified:

In this case the perturbation in \two ~is no
longer trivial and in fact it can be thought of as radius changing perturbation
in the following sense. Suppose $X^\mu$ is compactified on a circle of
radius $R^\mu$. Take $\lambda_{\mu\nu} = \lambda_{\mu}\,\eta_{\mu\nu}$
(no summation).The coordinate redefinition of the form \three~will then
change the radius from $R^\mu$ to $R^\mu\,{\sqrt {1+\lambda_{\mu}}}$
(no summation). Since the
spectrum of the theory changes with compactification radius, physics of
the perturbed action can not be reproduced by the original action by
target space coordinate redefinition. Hence we conclude that the marginal
deformations of the kind mentioned above are genuine perturbations for
compactified coordinates.

For later convenience, it is useful to discuss this situation in BRST
closed string field theory\SZ. In what follows, we will show that this
marginal perturbation can be understood as pure gauge deformation in
string field theory, if only $X^\mu$ is uncompactified and is not the case
otherwise. To linear order, the string field $|\Psi\rangle$ satisfies the
equation of motion
$$
Q_B |\Psi\rangle = 0.
\eqn\four
$$
The linearized gauge variation of the string field is
$$
\delta |\Psi\rangle = Q_B |\Lambda\rangle .
\eqn\five
$$
Here $Q_B$ is the nilpotent BRST charge of the first quantized theory and
$\Lambda$ is the gauge transformation parameter.\foot{We take the
off-shell string fields to be of ghost number two and they are
annihilated by $b_0^-$ and $L_0^-$. Similarly $\Lambda$ are the parameters
with ghost number one and as before are annihilated by $b_0^-$ and $L_0^-$.}

For our purpose we expand $|\Psi\rangle$ in component fields as
$$
|\Psi\rangle = h_{\mu\nu} c_1\bar c_1 \alpha_{-1}^\mu
\bar\alpha_{-1}^\nu |0\rangle + d (c_1c_{-1} -\bar c_1 \bar
c_{-1})|0\rangle + .......
\eqn\six
$$
Here $h_{\mu\nu}$ and $d$ are the string fields of tensorial rank two and
zero respectively and .... contains all other fields which are not
important for our present discussion. Now let us take a particular gauge
transformation
$$
|\Lambda\rangle ={lim}_{p\rightarrow 0}
\xi_{\mu\nu}(c_1\alpha_{-1}^\mu -\bar c_1\bar \alpha_{-1}^\mu){{e^{ipX^\nu}
-1\over ip}}|0\rangle .
\eqn\seven
$$
Notice that this is a valid gauge transformation parameter when $X^\mu$ is
uncompactified if we work with finite $p$ and take the limit at the end
of the calculations.

Upon acting $Q_B$, we get,
$$
Q_B |\Lambda\rangle = - 2i\xi^{\mu\nu}c_1\bar c_{1}\alpha_{-1}^\mu
\bar \alpha_{-1}^\nu |0\rangle - i {\xi_{\mu}}^{\mu}(c_1c_{-1} - \bar c_1
\bar c_{-1})|0\rangle.
\eqn\eight
$$
Now the gauge transformed string field in component form looks like
$$
|\Psi^\prime\rangle = (h_{\mu\nu} - 2i \xi_{\mu\nu}) c_1\bar c_1 \alpha_
{-1}^\mu
\bar\alpha_{-1}^\nu |0\rangle + (d - i{\xi_{\mu}}^\mu )
(c_1c_{-1} -\bar c_1 \bar
c_{-1})|0\rangle + .......
\eqn\nine
$$
Since $|\Psi^\prime\rangle$ and $|\Psi\rangle$ differ by a pure gauge
state, they are certainly equivalent string field cofigurations atleast to
linear order. On the other hand, adding \eight~ to string field
$|\Psi\rangle$ would mean adding $\int \lambda_{\mu\nu}\partial X^{\mu}
\bar \partial X^\mu$ in the corresponding sigma model ($X^\mu$ uncompactified)
action with the
identification $\lambda_{\mu\nu} = - 2i \xi_{\mu\nu}$ and with required change
in the dilaton coupling\VEN.

Now we turn to the case when $X^\mu$ is compactified. Notice that since in
this case the vertex operators take discrete momenta depending upon the
radius, we can not use the limiting procedure as in \seven. Instead, we
write $|\Lambda\rangle$ as
$$
|\Lambda\rangle =
\xi_{\mu\nu}(c_1\alpha_{-1}^\mu -\bar c_1\bar \alpha_{-1}^\mu)X^\nu
|0\rangle .
\eqn\ten
$$
This corresponds to a coordinate transformation parameter
$\xi_{\mu}(X)$ going as $\xi_{\mu\nu} X^\nu$, which is not a
well defined gauge transformation parameter. Hence
$Q_B |\Lambda\rangle$ should not be
considered as a gauge deformation. Consequently, $|\Psi\rangle$ and
$|\Psi^\prime\rangle$ will then be two inequivalent string field
configurations, which amounts to saying that
the corresponding perturbation in the sigma-model
acts as a non-trivial deformation of the original theory.

In the next section we analyze the `new' states in $c = 1$ matter coupled to
Liouville theory. We show that since the Liouville direction is
uncompactified, situation is very similar to the first case. In
particular, deformations by those type of states can
be identified as pure gauge deformations in the corresponding string field
theory.
\bigskip
\endpage
\centerline {\bf 3. Analysis of `new' states in d=2 string theory}

The action for $d=2$ string at $SU(2)$ point is\foot{Throughout this
section we follow the notatations and conventions of \WZ}
$$
S = {1\over 8\pi}\int d^2x{\sqrt h}(h^{ij}\partial_i X \partial_j X +
h^{ij} \partial_i \phi\partial_j \phi ) - {1\over 2{\sqrt 2}\pi}\int
d^2x{\sqrt h}\phi R^{(2)}.
\eqn\eleven
$$
Here $h$ and $R^{(2)}$ are the world sheet metric and Ricci scalar
respectively and $\phi$ is the Liouville coordinate with background charge
$2{\sqrt 2}$.

\noindent{\bf 3.1. States of Ghost number two:}

Among the discrete states, annihilated by $Q_B$ and $b_0^-$, at ghost
number two, there are states of the form:
$$
(a + \bar a) Y_{s,n}^+\bar O_{s-1,n^\prime};~~(a + \bar a) O_{s-1,n} \bar
Y_{s,n^\prime}^+
$$
where
$$
Y_{s,n}^+ = cV_{s,n}e^{{\sqrt 2}(1-s)\phi}
$$
with  $s = 0, 1/2, 1,....; n = s, s-1, ....,-s$ and $V_{s,n}$ is a primary
field constructed from the matter sector $X$. On the other hand, $O_{s,n}$
are the ground ring operators of ghost number zero such that
$$
O_{s,n} = x^{s+n}y^{s-n}
$$
with
$$\eqalign{
&x = O_{{1\over 2},{1\over 2}} = (cb + {i\over {\sqrt 2}}(\partial X - i
\partial \phi)) e^{i(X + i \phi)/\sqrt 2}\cr
&y = O_{{1\over 2},{-1\over 2}} = (cb - {i\over {\sqrt 2}}(\partial X + i
\partial \phi)) e^{-i(X - i \phi)/\sqrt 2}\cr}
$$
The operator $a$ which plays a crucial role in constructing these new
moduli is
$$
a = [Q_B^L,\phi] = c\partial \phi + {\sqrt 2}\partial c.
$$
where $Q_B^L$ is the holomorphic part of the BRST charge.

The first non-trivial example of these kind of new operators are states
with $s = 1, n = n^\prime = 0$. We take the combination
$$
\eqalign{
&(a + \bar a) Y_{1,0}^+\bar O_{0,0 } - (a + \bar a) O_{0,0} \bar
Y_{1,0}^+\cr
&= -c\bar c(\partial X\bar \partial \phi + \partial
\phi\bar\partial X) - {\sqrt 2}(c\partial c\partial X + c\bar\partial\bar
c\partial X - \bar c\partial c\bar\partial X - \bar c\bar\partial\bar
c\bar\partial X)\cr}
\eqn\twelve
$$
The 2-form corresponding to this state
that can be added to the sigma-model action is $(\partial X\bar\partial
\phi + \partial \phi\bar\partial X)$. This comes from the first term in
the right hand side of \twelve ~using descent equation. As we
will see soon the second term in r.h.s. of \twelve ~corresponds to
auxiliary field, which is not needed to be added to the action.
Hence the action now takes the form
$$\eqalign{
S_{pert} =& {1\over 8\pi}\int d^2 z (\partial X\bar \partial X + \partial
\phi\bar \partial \phi) -{1\over 2{\sqrt 2}\pi}\int R\phi d^2 z\cr
&+ \lambda\int
(\partial X\bar \partial \phi + \partial \phi\bar\partial X)\cr}
\eqn\thirteen
$$
Here $\lambda$ is the infinitesimal coupling assciated with the new state.
Clearly, the action \thirteen ~is not coformally invariant since the total
central charge differs from 26 by order ${\cal{O}}(\lambda^2)$. To
preserve the conformal invariance to this order, we need to add a term in the
ac
   tion of
the form ${\lambda^2\over 4\pi{\sqrt 2}}\int R\phi$.
With such changes we get,
$$
\eqalign{
S_{pert} =& {1\over 8\pi}\int d^2 z (\partial X\bar \partial X + \partial
\phi\bar \partial \phi) - {{\sqrt{1-\lambda^2}}\over 2
{\sqrt 2}\pi}\int R\phi d^2 z\cr
&+ \lambda\int
(\partial X\bar \partial \phi + \partial \phi\bar\partial X).\cr}
\eqn\fourteen
$$

We now redefine the target space coordinates as follows:
$$
X^\prime = X + \lambda \phi;~~\phi^\prime = {\sqrt {1- \lambda^2}}\phi .
\eqn\fifteen
$$
Hence the perturbed action ~in terms of these coordinates can be written as,
$$
S_{pert} = {1\over 8\pi}\int d^2x{\sqrt h}(h^{ij}\partial_i X^\prime \partial_j
X^\prime  +
h^{ij} \partial_i \phi^\prime \partial_j \phi^\prime
 ) - {1\over 2{\sqrt 2}\pi}\int
d^2x{\sqrt h}\phi^\prime R^{(2)}.
\eqn\sixteen
$$
This is exactly same as \eleven ~with $\phi$ and $X$ replaced by
$\phi^\prime$ and $X^\prime$ respectively. Notice that since the Liouville
direction is uncompactified, the coordinate choice \fifteen ~is globally
possible. We therefore conclude that this new moduli will act as trivial
perturbation to the original theory. The situation is quite similar to the
case of marginal deformation in critical string theory as discussed before.

Now we pass on to the corresponding BRST string field theory where we
identify this deformation to be a pure gauge deformation with singular
gauge transformation parameter. We expand the string field $|\Psi\rangle$
as
$$
|\Psi\rangle = \tilde g ~c_1\bar c_1 (\alpha_{-1}\bar \phi_{-1} +
\phi_{-1} \bar \alpha_{-1})|0\rangle + i \tilde s ~c_0^+(c_1\alpha_{-1} - \bar
c_{1}\bar \alpha_{-1})|0\rangle + .....
\eqn\seventeen
$$
Here $\tilde g$ and $\tilde s$ are the component string fields and
$\alpha_n$, $\phi_n$ are the matter and Liouville oscillators
respectively. We choose the gauge transformation parameter to be
$$
|\Lambda\rangle = lim_{p\rightarrow 0} i \lambda (\bar c_1 \bar \alpha_{-1}
- c_1 \alpha_{-1}){e^{p\phi} -1\over p}|0\rangle .
\eqn\eighteen
$$
With this choice of $|\Lambda\rangle$, we find,
$$
Q_B |\Lambda\rangle = \lambda c_1\bar c_1 (\alpha_{-1}\bar \phi_{-1}
+ \bar \alpha_{-1}\phi_{-1}) |0\rangle
- 2 i \lambda c_0^+(c_1\alpha_{-1} - \bar c_1 \bar
\alpha_{-1})|0\rangle
\eqn\nineteen
$$
Adding this to the string field configuration \seventeen, we get the
changes in the component fields as
$$
\delta \tilde g = \lambda~; ~~~\delta\tilde s = - 2\lambda.
\eqn\twenty
$$
If we redefine the fields as
$$
g = \tilde g~;~~~s = \tilde s + 2 \tilde g,
\eqn\twentyone
$$
then the corresponding transformations are
$$
\delta g = \lambda~;~~~\delta s = 0.
$$
Since, to linear order, the string field theory equation of motion is $
Q_B |\Psi\rangle = 0$, it is easy to see that the  equation of motion
involving $s$ is: $s = 0$, showing that it is an auxiliary field to this
order and can be set to zero using its equation of motion.

In sigma model language, this would correspond to adding a term
$\lambda\int (\partial \phi \bar\partial X + \bar\partial\phi\partial X)$
to the action atleast to lowest order in $\lambda$. Recall that in
order to preserve the conformal invariance, we had to add higher order
$\lambda$ dependent curvature term in the action. In string field theory,
we believe, it will be taken care of automatically. There are higher
order $\lambda$ dependent corrections to gauge transformation in \five.
In particular, the $\lambda^2$ order correections will have non-zero
contribution along the dilaton direction, hence will guarantee the
conformal invariance to this order.

What we have noticed so far can be generalized to any of the members of
this kind in ghost number two sector.
Though it is hard to obtain explicit target space redefinitions for a
general state in that tower, nevertheless, it is easy to understand their
effect from string field theory. Since a general member of the series can
be written as
$$
( a + \bar a ) Y_{s,n}^+ \bar O_{s-1, n^\prime} = Q_B(\phi Y_{s,n}^+ \bar
O_{s-1, n^\prime}),
\eqn\twentytwo
$$
following our earlier argument, the effect of \twentytwo ~can be thought
of as linearized pure gauge deformations when added to the string field
configuration. As a result, the two classical solutions differing
by \twentytwo
{}~should be regarded as two equivalent solutions.
Note that in the case where $X$ is a noncompact scalar field many more
marginal operators become pure gauge in this sense. For example in the
first multiplet for $s = 1, n = n' = 0$, the operator
$\partial X\,\bar\partial X$ gave radius changing perturbation in
the compact case.
In the noncompact case, this is pure gauge. This is true for $Y_{s, n}^+
\bar Y_{s, n}^+$ in general. It is nontrivial for compact $X$ only.

Given two world sheet actions differing from each other by a marginal
perturbation, one knows \AS  how to construct solutions of equation of
motion in corresponding string field theory. If the perturbation is
non-trivial (i.e. can not be removed by suitable coordinate redefinition),
the two solutions will certainly be inequivalent since the starting actions
are different. Now, suppose we know the moduli corresponding to
\twentytwo ~that can be added as a marginal perturbation to the sigma
model action. In the sense of above argument it would rather correspond to
trivial deformation of the theory because  the classical solutions of
original and perturbed actions are equivalent.
Since the new moduli will have higher
tensorial rank, the coupling will also be the same.\foot{Here we use
$X^\mu$ as both matter and Liouville  coordinates. Hence, in this
notation, the metric $G_{\mu\nu}$ is of tensorial rank two.} Thus in
general in the original action we need to keep higher non-zero background
couplings. Moreover, the target space redefinition will no longer be the
same as \fifteen. It will include derivative dependence among the
coordinates. Given a particular gauge transformation in the string field
theory, there is no straightforward way to see the corresponding target
space redefinition in sigma-model,
but, a rather indirect method of identification has
been discussed in \MM.

\noindent{\bf 3.2. States of Ghost Number One}

Here we discuss about the discrete states with ghost number one satisfying
$(b_0
- \bar b_0)$ condition. Notice that these states can be thought of as gauge
transfomation parameters in string field theory. As we mentioned before,
the gauge transformation parameters in string field theory are of ghost
number one and are annihilated by $(b_0 -
\bar b_0)$. Under infinitesimal gauge transformation, string field
changes as
$$
\delta (\Psi) = Q_B (\Lambda) + g [\Psi\Lambda] + {\cal
{O}}(\Psi^2)
\eqn\twentythree
$$
If $\Lambda$ happens to be a BRST closed state, then the first term in
the right hand side drops out. This is the case for global transformation,
where fields transform homogeneously. More over, if $\Lambda$ is
BRST exact, i.e.
$$
|\Lambda\rangle = Q_B |\Theta\rangle
\eqn\twentyfour
$$
then
$$
\delta (|\Psi\rangle) = g[\Psi Q_B\Theta] + {\cal{O}}(\Psi^2).
\eqn\twentyfive
$$
Now, if $|\Psi\rangle$ satisfies the equation of motion, then,
$$
\delta ( |\Psi\rangle) = Q_B g[\Psi \Theta] + {\cal{O}}(\Psi^2).
\eqn\twentysix
$$
If we allow $[\Psi\Theta]$ to be a valid gauge transformation parameter,
then $|\Psi\rangle$ and $|\Psi\rangle$ + $\delta |\Psi\rangle$
are equivalent solutions.
So the nontrivial
global part of the gauge symmetries are those for which $\Lambda$
belongs to BRST cohomology.
This is precisely the situation that occurs in $d=2$ string theory. Among
all the ghost number one states, some are of the form $( a + \bar a )
O_{u,n}\bar O_{u,n^\prime}$ with
$$
( a + \bar a )
O_{u,n}\bar O_{u,n^\prime} = Q_B (\phi O_{u,n}\bar O_{u,n^\prime})
\eqn\twentyseven
$$
Since, we allow the states $\phi O\bar O$ (in the limiting sense as
defined in eqn. \eighteen) in the string field configuration,
we can identify $\Theta$ with these states. As a consequence, $\Lambda =
Q_B\Theta$ should be considered as trivial global symmetry generator
in this string field theory.

For the first multiplet,
$\Theta = \phi$; $\Lambda = a + \bar a$. This corresponds to a constant
gauge parameter for the antisymmetric tensor field. The corresponding
charge measures the Liouville winding mode. This acts trivially on
all the states. The charges from the higher multiplets do not act
trivially but transform ghost number two states into the new ghost number
two states as we saw.
\bigskip
\centerline{\bf 4. Comments on higher ghost number states.}

We could take the cohomology at higher ghost numbers and apply the same
argument. For example, ghost number three states can be interpreted as
antifield modes. If we take free string field action $S_0$ and carry out
Batalin-Vilkovisky quantization procedure for the action, we would get
$S = \langle \Psi| c_0^-Q_B|\Psi\rangle$ with $|\Psi\rangle$ having
components from {\it { all}} ghost numbers, with $L_0^-|\Psi\rangle$ =
$b_0^- |\Psi\rangle$ = $0$. \foot{ The original action $S_0$ is obtained
by restricting $|\Psi\rangle$ to ghost number two.} The components from
ghost number one would correspond to target space ghost and those from
ghost number three would correspond to target space antifields and so on.

Note that this action has a gauge invariance $\delta |\Psi\rangle =
Q_B|\Lambda\rangle$, where $L_0^-|\Lambda\rangle = b_0^-\Lambda\rangle =
0$, $|\Lambda\rangle$ has components from all ghost numbers.
Now if we want to solve the equations of motion comming from $S$ and get
solutions upto this gauge transformation, we end up getting the BRST
cohomology for all ghost numbers.

For a generic action of the form
$$
S_0 = \Psi^i A_{ij} \Psi^j,
\eqn\twentyeight
$$
we have
gauge invariance of the form $\delta \Psi^i = {R_\alpha}^i\Lambda^\alpha$,
if $A_{ij}R_\alpha^i = 0$. Corresponding Batalin-Vilkovisky action ~\MH
would
then be
$$
S = \Psi^i A_{ij} \Psi^j + \Psi_i^* R_\alpha^i C^\alpha +.....
\eqn\twentynine
$$
where $C^\alpha$ are the target space ghosts and $\Psi^*$
are the antifields. The
equations of motion of the antifields are $\Psi_i^*R_\alpha^i = 0$. If the
action ended with these two terms, the gauge transfomations would have been
$$
\delta \Psi^i = R^i_\alpha \Lambda^\alpha,
{}~~\delta\Psi_i^* = A_{ij}\chi^j,~~\delta C^\alpha = 0.
\eqn\thirty
$$
Hence the gauge variation of the antifields would have corresponded to
off-shell field configuration. Its equations of motion makes it
`orthogonal' to the gauge transformations of the fields (since
$\Psi_i^*\delta \Psi^i = 0$). Hence the independent antifields solutions,
modulo gauge transformations, would have been in one-to-one correspondence
with the genuinely independent physical solutions.

In string field theory, the quadratic action does go beyond the first two
terms. However, as long as we stick to the space of conformal fields, the
conclusion presented here holds. Namely there is a one-to-one correspondence
between fields and antifields, pure gauge and off-shell states. However as
one comes out of the space of conformal fields, these correspondances
break down.
Consider a ghost number two state of the form $Q_B \phi|\Lambda\rangle$,
where $|\Lambda\rangle$ is a member of the ghost number one cohomology.
It is a pure gauge field mode for a gauge parameter going as $\phi$.
This has a nontrivial overlap with some states in the ghost number three
BRST cohomology. From the target space point of view, this is caused by
the presence of the boundary term when we do integration by parts in an
expression of the form $\int \Psi^*R\Lambda$, $R$ being generically some
differential operator. Since equations for the antifields are
$R^\dagger\Psi^* = 0$, we will get a boundary contribution after
integrating by parts when
$\Lambda ~\sim ~\phi$. Similarly, pure gauge antifield modes will have
overlap with physical fields.

Hence if we want the conjugates of the fields with ghost number less than
three, which is not of the form $Q_B(\phi{\cal{O}})$, we might have to
keep the states of the form $Q_B(\phi {\cal{O}})$ for the ghost numbers
greater than or equal to three. In fact, it turns out only those are the
ones that are to be kept. The complete meaning of this is not clear to us.

\noindent{\bf{Acknowledgements:}} We have greatly benefitted from
discussions with S. Mukhi, S. K. Rama, A. Sen and S. R. Wadia.

\refout

\end